\documentstyle[12pt]{article}

\input{tcilatex}

\begin{document}

\title{Amplification of gravitational \ waves during inflation in Brans-Dicke
Theory (Revised Version)}
\author{Marcelo S.Berman$^{(1)}$ \and and Luis A.Trevisan$^{(2)}$ \\
(1) Tecpar-Grupo de Projetos Especiais.\\
R.Prof Algacir M. Mader 3775-CEP 81350-010\\
Curitiba-PR-Brazil\\
Email: marsambe@tecpar.br\\
(2) Universidade Estadual de Ponta Grossa,\\
Demat, CEP 84010-330, Ponta Grossa,Pr,\\
Brazil \ email: latrevis@uepg.br}
\maketitle

\begin{abstract}
Due to a previous result [3], there is a possibility of exponential
inflationary phase in Brans-Dicke theory, even for an equation of state $%
p=\alpha \rho $ with $\alpha >0.$ In this case , we show that the
perturbation in the metric would lead to an enormous amplification of
gravitational waves that could detectable, making it possible to discard
General Relativity theory in favour of Brans-Dicke theory.

PACS \ 98.80 Hw
\end{abstract}

\newpage

\begin{center}
\bigskip {\bf {AMPLIFICATION OF GRAVITATIONAL WAVES DURING INFLATION IN
BRANS-DICKE THEORY.(Revised Version)} }

{\bf {MARCELO S. BERMAN and LUIS A. TREVISAN.} }
\end{center}

\section{Introduction}

\noindent One of the possibilities for inflationary models is the
exponential scale-factor \cite{1}\cite{2}\cite{4}. In such case, Einstein's
theory of General Relativity yields the equation of state:

\[
p=-\rho 
\]
where $p,$ $\rho $ \ stand for \ pressure and energy density. Berman and Som 
\cite{9} \ showed that under this equation of state, all the conformally
flat space-times render a constant expansion rate in G.R., but the
literature on Brans-Dicke [11] cosmology shows the possibility for other
equations of state[3], like $p=\alpha \rho $ with $\alpha >0.$ Being a
viable alternative, B.D. gravitational theory must be taken into account
when one delves into constant expansion rates. ``Extended '' inflation, put
forward by La and Steinhardt ,\cite{10} ,with support of a paper by Barrow
and Maeda \cite{12}, is one of such attempts. It seems to us that the
current experimental data are not capable of outruling B.D. theory ( or its
generalization involving a variable coupling ``constant '').

On the other hand, the study of gravitational waves production in the
expanding Universe has been a subject of major research projects, either on
theoretical or in the experimental fields. We shall show that, in contrast
with G.R., we would have an enormous amplification of gravitational waves in
the inflationary phase with $p=\alpha \rho $ and $\alpha >0$, in Brans-Dicke
theory.

\section{Derivation of Amplification for G.W 's.}

Barrow et al. \cite{1} obtained the gravitational waves equation for a flat
Robertson-Walker's metric, in scalar tensor theories of gravity, as
considered by Barrow \cite{2} Some particular cases were solved \cite{1} ,
in the conformal time $\eta $ system, defined by

\begin{equation}
dt=ad\eta
\end{equation}

where $a=a(t)$ stands for the scale-factor in the metric:

\begin{equation}
ds^{2}=dt^{2}-a^{2}(t)\left[ dx^{2}+dy^{2}+dz^{2}\right]
\end{equation}

For an account of Brans-Dicke cosmologies, we refer to Weinberg \cite{8}

The amplification in Brans-Dicke theory was derived by Barrow et al.\cite{1}
, and is essentially given by the $Y_{k}(t),$ with the equation:

\begin{equation}
\ddot{Y_{k}}+\left[ \frac{\dot{\phi}}{\phi }-\frac{\dot{a}}{a}\right]\dot{%
Y_{k}}+\left[ \frac{k^{2}}{a^{2}}-\frac{2\ddot{a}}{a}-2\frac{\dot{a}\dot{\phi%
}}{a\phi }\right] Y_{k}=0
\end{equation}
where $k\equiv |\overrightarrow{k}|$ , the \ comoving wave vector,

\begin{equation}
k=\frac{2\pi a}{\lambda }
\end{equation}
and $Y_{k}$ is \ the \ amplitude. The perturbation in the metric is given by
:

\begin{equation}
h_{k}=\frac{Y_{k}}{a^{2}}
\end{equation}

For a B.D inflationary phase of the exponential type, given by:

\begin{equation}
a=a_{0}e^{Ht}
\end{equation}
($a_{0,}H$ constants), we find,

\begin{equation}
\phi =\phi _{0}e^{\beta t}
\end{equation}
($\beta =const).$

\begin{equation}
\rho =\rho _{0}e^{\gamma t}
\end{equation}
($\gamma ,\rho _{0}$ constants), \ and,

\begin{equation}
p=\alpha \rho
\end{equation}
($\alpha =const)$.

Relation (9) makes for a perfect gas equation of state, while the relations
between the constants of the theory were found by Berman and Som \cite{3} .
In particular,

\begin{equation}
\gamma =-3H(1+\alpha )=\beta 
\end{equation}
When we plug (6), (7), (8) and (9) into (3), we find an exponentially
increasing solution for $h_{k}$, so that there is indeed amplification of
gravitational waves, in Brans-Dicke theory, for such a phase, but only for
the case $\alpha >0.$ This can be seen from the equation for \ $Y_{k,}$
which is,

\begin{equation}
\ddot{Y_{k}}-H\left[ 4+3\alpha \right] \dot{Y_{k}}+\left\{ 2H^{2}\left[
2+3\alpha \right] +\frac{k^{2}}{a_{0}^{2}}e^{-2Ht}\right\} Y_{k}=0.
\end{equation}
When we neglect the term in $e^{-2Ht}$ , we find,

\begin{equation}
Y_{k}=Be^{\delta t}
\end{equation}
where $B,C$ and $\delta $ are constants, as a solution where $\delta =\delta
_{1}=2H$ or $\delta =\delta _{2}=H(2+3\alpha )$ $.$The perturbation in the
metric is given by a constant $h_{k}$\ for the first solution. In the second
case we find

\begin{equation}
h_{k}=Ce^{3\alpha Ht}
\end{equation}
so that, for $\alpha >0$ an extraordinary amount of amplification is
possible. The condition for $\alpha $ results from requirement that $h_{k}$
increase with time in expression (13).

Incidentally, Berman and Som[3] found the condition to be obeyed by $H:$

\begin{equation}
H^{2}=\frac{8\pi \rho _{0}}{3\phi _{0}\left[ -2-3\alpha -\frac{3}{2}%
(1+\alpha )^{2}\right] }>0
\end{equation}
In reference \cite{3} , Berman and Som also found

\begin{equation}
\omega =\frac{4}{\left( 1+\alpha \right) }\left[ \frac{9\alpha ^{2}+12\alpha
+1}{9\alpha ^{2}-30\alpha -3}\right]
\end{equation}
with $\alpha \neq -1.$

The case $\alpha =-1$ recovers G.R, because $\omega \rightarrow \infty .$
The relation between $G$ (Newton's \ gravitational \ constant) and $\phi ,$%
is \cite{8}:

\begin{equation}
G=\left[ \frac{2\omega +4}{2\omega +3}\right] \phi ^{-1}>0
\end{equation}

\bigskip

\section{Conclusion}

In conclusion, we have shown, by considering the inflationary (exponential)
phase, in B.D theory, that amplification of gravitational waves is possible,
with an enormous output in the $\alpha >0$ case. This was a result obtained
during the inflationary phase, and it does not need any phase transition
consideration. In contrast, the corresponding equation in G.R, obtained by
Grischuk (5),(6),(7), presents no such\ large amplification, for exponential
scale-factor, when we focus only on such phase , because it is found $%
h_{k}=const$ .

It turns out that the gravitational waves originated in such period can be a
decisive test for B.D. theory in particular, and to the existence of an
inflationary period in the past history of the Universe. It also may be a
test for the the homogeneous and isotropic metric that represents, according
to our present model, the Universe, provided that we try to measure the same
properties (homogeneity and isotropy) in the spectrum of such gravitational
waves. The detection of such g.w`s could allow us to outrule General
Relativity theory in favour of other scalar-tensor theory.

\medskip

\noindent {\bf Acknowledgments}

Both authors thank support by Prof Ramiro Wahrhaftig, Secretary of Science
,Technology and Higher Education of the State of Paran\'{a}, and by our
Institutions, especially to Jorge\ L.Valgas, Roberto Merhy, Mauro K.
Nagashima, Carlos Fior, C.R. Kloss, J.L.Buso, and Roberto Almeida. The
report of an anonymous refeere saved us from a ''faux pas'' , so that to him
we extend our thanks.

\end{document}